# Information, communication and music: Recognition of musical dissonance and consonance in a simple reservoir computing system

*Dawid Przyczyna,[a,b] Maria Szaciłowska,[a] Marek Przybylski,[a,b] Marcin Strzelecki, Konrad Szaciłowski[a]\**

[a] AGH University of Science and Technology, Academic Centre for Materials and Nanotechnology, al. Mickiewicza 30, 30-059 Kraków, Poland
[b] AGH University of Science and Technology, Faculty of Physics and Applied Computer Science, al. Mickiewicza 30, 30-059 Kraków, Poland
[c] Faculty of Composition, Interpretation and Musical Education, Academy of Music in Kraków, ul. Św. Tomasza 43, 31-027 Kraków, Poland

\* corresponding author, szacilow@agh.edu.pl

**Abstract**

Reservoir computing is an emerging, but very successful approach towards processing and classification of various signals. It can be described as a model of a transient computation, where influence of input changes internal dynamics of chosen computational reservoir. Trajectory of these changes represents computation performed by the system. The selection of a suitable computational substrate capable of non-linear response and rich internal dynamics ensures the implementation of simple readout protocols. Signal evolution based on the rich dynamics of the reservoir layer helps to emphasize differences between given signals thus enabling their easier classification. Here we present a simple reservoir computing system (single node echo-state machine) implemented on Multisim platform as a tool for classification of musical intervals according to their consonant or dissonant character. The result of this classification closely resembled sensory dissonance curve, with some significant differences. A deeper analysis of the received signals indicates the geometric relationships between the consonant and dissonant intervals, enabling their classification.

## 1. Introduction

Communication between organisms is an ubiquitous phenomenon, both at intraspecies and interspecies level in all kingdoms: *Archaea* [1], *Bacteria* [2,3], *Protista* [4], *Fungi* [5], *Plantae* [6] and *Animalia* [7]. Surprisingly, primitive communication was detected even between individual virions [8]. All these organisms possess both intracellular, intraorganismic and transorganismic communication protocols, however the most complex and interesting ones, from the point of view of information theory, are those between individual organisms. In most cases the intracellular/intraorganismic communication is based on signaling molecules, the same concerns most of the interorganismic and interspecies communication protocols. Communication in general can be described as a sign-mediated interaction between at least two living entities, which share the common repertoire of signs representing a form of natural language. These signs may be combined according to syntactic rules in various contexts (according to pragmatic rules) and used to transport biologically relevant information. Almost all kingdoms of life use molecules as the only available communication tool, whereas animals add vocal and visual communication tools to their repertoire of available signs. In humans these evolutionary novelties dominate, almost completely, over the molecular language, however "molecular senses" of olfaction and gustation are still significantly important. Most of animals use senses of vision and hearing for most of their communication purposes. Whereas our own (human) senses seem to be impaired (as compared with some predatory birds), their ability to process signals is still amazing. We have also developed unique ways of communication: music and language, manifested sonically as speech, and graphically as writing. These tools provide an unprecedented opportunity to communicate language and emotions using graphical symbols and aesthetic, religious and cultural feelings via organized sounds of different parameters like pitch, durations, and timbral qualities, arranged in melodic, rhythmic, and harmonic (tonal) patterns.

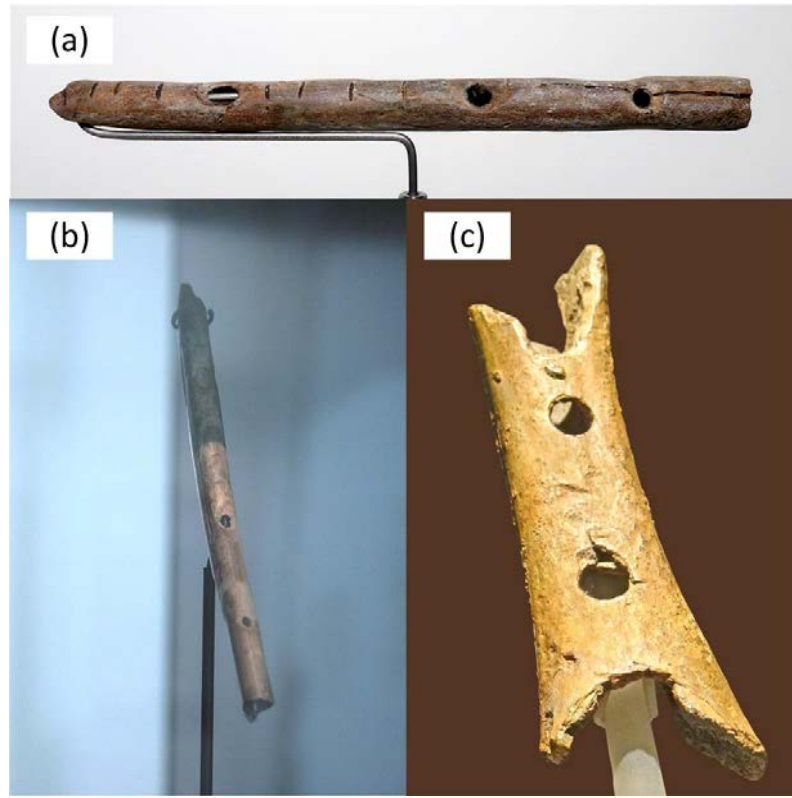

*Figure 1. Paleolithic musical instruments: upper Paleolithic from Geissenklösterle (a), middle Paleolithic flute, ca. 35000-40000 years, (b) and neandertalic flute from Divje Babe Cave (Slovenia, 55000 years ago, c). Photos courtesy José-Manuel Benito (a), Marco Ciamella (b) and Jean-Pierre Dalbéra (c).*

Music is the only form of natural communication, that is created and perceived only by humans (however studies on animals indicate some aspects of sensitivity to music) [9,10]. Music belongs to human universals, *i.e.* elements, patterns, features, or notions that are common to all human cultures worldwide [11], however, according to some opinions, it does not convey any biologically-relevant information [12]. According to Guerino Mazzola "*music embodies meaningful communication and mediates physically between its emotional and symbolic layers*" [13]. The importance of music is exemplified by the discovery of Paleolithic musical instruments. Whereas most probably music at early times had no direct effect on the economy or a reproductive success, it may have had provided medium of social integration [14] (Fig. 1). As of today, the influence of *muzak* [15] on our decisions in supermarkets and retail centers proves its impact on real profits from these businesses. Nowadays music is one of the most ubiquitous human activity independently on any social and cultural attributes or intellectual abilities.

Music and speech are created and processed by distinctively various neural structures, but they have some common features: they are means of communication, have specific syntax

– *i.e.* there exists a set of rules defining proper combination of elements (words or notes) [16]. Some kinds of music, like European tonal music, have more strict syntax [17], some others (like dodecaphonic music) may be strictly organized while at the same time lacking of audible regularities [18]. Finally, there exist also genres of free improvised, experimental music, and anti-music movements, which aims at breaking traditional regularities [19]. Such exceptions and declared negation of musical syntax also confirms the existence of one. Music is a domain of human artistic and entertaining activity, but also a field of vigorous research. Likewise information, music is a notion very difficult to define in precise terms. Dislike speech, music seems not usually meant for direct communication purposes, especially of biological importance [12]. Conversely, it is meant to trigger various emotional responses in recipients due to aesthetical feelings [20]. On the other hand, music is a very well organized structure. Even the denial of the existence of such structure, conceptually declared by the author, proves the existence of specific 'musical language" with appropriate grammar, syntax and vocabulary – the harmony, rhythmical patterns, timbres and their mutual relations [21-23].

Therefore, not every combination of sounds should be considered as music, and specific fractal signatures can be assigned to specific genres [24,25]. The simplest musical message, melody, can be defined as appropriate time sequence of quantized frequencies (Fig. 2). These frequencies are called steps in musical scale. Most musical systems are founded on a concept of the octave: an interval between frequencies of $f$ and $2f$. Octave is an interval between the first and second harmonics of the harmonic series. Therefore octave is considered as a natural phenomenon that has been referred to as the "*basic miracle of music*", the use of which is "*common in most musical systems*"[26].

In the music of European origin, an octave is divided into 12 steps (evenly spaced the case of equal temperament, but unevenly spaced in the case of natural tuning), called semitones. Also, there exist many different tuning systems [27], and octave divisions (like Balinese and Javanese gamelan systems). Other musical systems, both traditional (e.g. the Middle East, India and Far East) as well as modern experimental musical genres, use different intervals, including division of octave into 4, 5, 7, 34 (to name only a few possibilities) or even 96 equal steps, leading to the whole musical tuning continuum [28,29]. A characteristic feature of European music is the specific concept of musical harmony, which originates from geometrical foundations [30,13] and may be considered as a key component of theory and practice.

Musical harmony is a complex notion related to the perception of individual and superimposed sounds. The notion of musical harmony includes: (i) the pure content of the set of frequencies heard at given time (including their harmonic components responsible for a timbre of an individual note), (ii) mutual relation of a set of simultaneously played notes (i.e. the verticality of the chord), (iii) the tonal context and context of adjacent chords which determine the quality (called harmonic function) of particular chord [31], and (iv) the position and relation of a chord in relation to the melody at given moment [32,33].

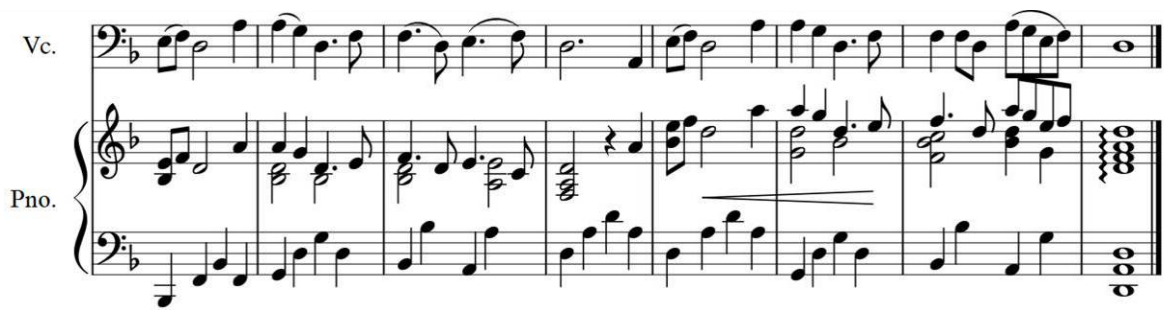

*Figure 2. An example of a musical score: final bars of 'Rains of Castamere' by R. Djawadi (arranged for cello and piano by M. Magatagan).*

The two fundamental notions associated with musical harmony are consonance and dissonance [34,35]. This notion concerns the aesthetic feeling evoked by two (or more) sounds played simultaneously. The first detailed study on dissonance and consonance comes from works of Pythagoras of Samos, who related consonant and dissonant combination of tones with the ratio of the lengths of strings. Octave, perfect fifth and perfect fourth were recognized as consonant intervals and founded the basis in Pythagorean philosophy and cosmology. Further works of physicists and music theorists (including Gioseffo Zarlino, Vincenzo and Galileo Galilei, Marin Mersenne, René Descartes, Daniel Bernoulli, and others) came to conclusion that the feeling of consonant and dissonant combinations of sounds relied on the ration of frequencies. Further works of French composer Jean-Philippe Rameau associated the ratios of harmonic overtones with musical intervals and concluded that harmonic series are foundations of musical harmonies. It was further developed by Jean Rond d'Alembert, Leonhard Euler and Jean Baptiste Joseph Fourier, who provided a complete description of harmonic series for string, air columns and other physical systems. Later on Hermann von Helmholtz has focused on the sense of audition and have developed theory which is based on interaction of acoustic waves in human ear.

Dissonance was associated with sound 'roughness' and 'beating' being a result of interference of acoustic waves, especially when two tones are of similar frequencies [36]. This theory has been falsified by recent experiments reported by Mc Dermott et al. [37]. It was found that the feeling of dissonance and consonance can be easily separated from the feeling of sound roughness. Furthermore, dissonance and consonance can be easily distinguished when sounds are played diotically (two tones to the ears) or dichotically (one tone to one ear). This, however does not falsify the role of harmonic components in the sound, as discussed by Plomp & Levelt [38] and recently by Sethares [33,39], but may indicate other physical background of dissonance perception. The approaches mentioned above state, that the tone combination is dissonant if there are dissonant ratios between higher harmonics of two tones and the degree of dissonance depends on the number of dissonant ratios. Upon proper assignment of parameters, the model gives very reliable results, comparable with the auditory evaluation of dissonance (Fig. 3a).

The dissonance function between two sine waves of frequencies $f_1$ and $f_2$ and corresponding amplitudes of $l_1$ and $l_2$ is defines as (1):

$$d(f_1, f_2, l_1, l_2) = l_{12}\left(c_1 e^{-b_1 s(f_2-f_1)} + c_2 e^{-b_2 s(f_2-f_1)}\right), \tag{1}$$

where

$$l_{12} = \min(l_1, l_2), \tag{2}$$

$$s = \frac{x^*}{s_1 f_1 + s_2} \tag{3}$$

and the parameters have the following values: $b_1$ = -3.51, $b_2$ = -5.75, $c_1$ = 5.00, $c_2$ = -5.00, $s_1$ = 0.0207, $s_2$ = 18.96 and $x^*$ = 0.24 [33]. The dissonance of complex sounds being a collection of sine wave components $f_0 \ldots f_n$ ($f_k = kf_0$) of amplitudes $l_1 \ldots l_n$ can be thus calculated as (4):

$$D = \sum_{i=1}^{n}\sum_{j=1}^{n} d(f_i, f_j, l_i, l_j) \tag{4}$$

The results of these calculations are presented in Fig. 3b-c. It should be noted, that this approach relies on the presence of harmonic components in each tone: for single sine signals only unison can be detected, addition of higher number of components results in isolation of octave (2 components), fifth (3 components) and fourth (4 components).

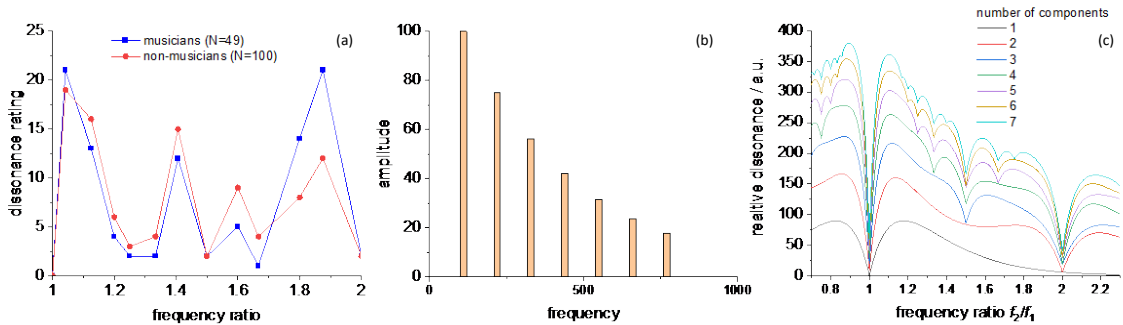

*Figure 3. Perception of dissonance by human subjects with (49 persons) and without (100 persons) musical training, based on data from Ref. [40] (a). Musical dissonance calculated for a base tone $A_2 = 110$ Hz and variable number of harmonic components: Fourier spectrum of the tone (b) and the calculated sensory dissonance curves for variable number of harmonic components (c).*

Despite the well-established musical theories [32] the automated classification of intervals, chords and clusters and recognition of consonance and dissonance has not been achieved in any system without prior training on the basis on the theory of harmony. Furthermore, the understanding of physical nature of dissonance and consonance is still not fully established [12], but it seems that it may have a background hidden deeply in neuronal dynamics [41,42]. It was found in a series of independent studies that infants prefer consonant over dissonant sound combinations [43]. Some primates also show the preference of musical consonance [44]. All these considerations point out the fact that the basic understanding of the phenomenon of musical consonance and dissonance is still missing. The motions of consonance and dissonance are antagonistic, and the dissonance can be described as a lack of consonance, or vice versa, depending on personal preferences. For two tones of frequencies $f_1$ and $f_2$ the beating frequency ($f_2-f_1$) results in sensation of 'roughness' once its value falls within particular range [45]. This range is related to critical bandwidth – a phenomenon created by the cochlea [46], responsible for the psychoacoustic effect of masking of one tone with another one, so the listener cannot distinguish their frequencies, a phenomenon related to cognitive dissonance [47,48,38,49]. From all these studied one can conclude that most probably music, and especially the perception of musical consonance and dissonance may have physiological roots [50]. Although some cultural, stylistic and psychological factors lead to individual, subjective aesthetic estimation of those sonorities [51-53], the dissonance-consonance opposition, as physiological phenomenon, remains objective by nature. Indeed, recent electrophysiological studies on human subjects indicate significant correlation between neural activity in selected brain regions induced by perception of musical dissonance and consonance [54-57]. Even more strikingly, significant preference of musical consonance was observed in numerous animal studies, e.g. infant chimpanzee (*Pan troglodytes*) [58], maccacs (*Macaca fascicularis*) [59], tamarins (*Saguinus oedipus*) [60], chickens (*Gallus gallus*) [61] and Java sparrows (*Padda oryzivora*) [62], determined either in behavioural or electrophysiological studies using implanted electrodes. All the studies mentioned above indicate an important biological component to the perception of music and the special role of musical consonance. They also underline the significance of music at very basic neurophysiological level in birds, mammals and humans. Therefore, we intend to check, if a simple neuromimetic device operating in unsupervised learning mode will be able to differentiate combinations of tones into consonant and dissonant categories. If perception of music is a neurophysiological process, then a device that mimics the dynamics of brain structures should show the same ability.

      The search for a system capable of advanced signal processing has turned authors attention towards reservoir computing and especially echo state machines, a subclass of

reservoir computing systems with delayed feedback. These systems are usually reported as mimicking the neural dynamics, furthermore, they can be relatively easy implemented both in hardware and in software. Recent findings in the field of single node echo state machines (SNESM) show a possibility of discrimination signal and pulse amplitudes without prior training [63,64] as well as application of simple photoelectrochemical dynamic systems for the recognition of handwritten characters [65] and improvement of chemical sensor performance [66-68]. It suggests that dynamic systems with appropriate feedback, fading memory and internal dynamics should be capable of advanced signal classification. These properties fit well into the notion of Reservoir Computing (RC) paradigm, where richness of possible states and their appropriate sensitivity/fading after stimulation can be used as effective platform for information processing. After ensuring that given computational substrate (or readout process) [69] presents appropriate dynamic conditions, simple readout processes should suffice in probing concomitant reservoir states thus performing computation (as the evolution of reservoir internal state represents its computing capabilities). By incorporating suitable substrate – for example a simple neuromimetic device as in this work – in the SNESM reservoir system, its information processing capabilities can be enhanced. The interested reader is referred to the literature on the RC subject presenting a rigorous mathematical description [70-72,68] along with precise description of prerequisites for reservoir layer [73-76] as well as some practical examples [64,77,63,66].

The goal of this study is to verify the utility of reservoir computing approach towards unsupervised classification of musical intervals a series of numerical tests have been designed. A series of musical intervals from tones belonging to natural scale has been constructed and subjected processing by the memristive synapses in a feedback loop *in silico*. Two different post-processing readout operation have been designed and their performance is compared.

**Experimental**

A simple numerical model of memristor was taken from the paper and used without any modifications as shown in Table 1 [78]. Figure 4 shows an equivalent circuit of this memristor. Two current sources $G_{r1}$ and $G_{r2}$ which have opposed polarities and operate in such a way so that $G_{r2}$ is responsible for charging the capacitor, and $G_{r1}$ for discharging it. Their operation is controlled by the necessary step functions used to determine which source is active each time, according to the applied voltage. Moreover, the problem of limiting the boundaries of $r(V_M, t)$ is addressed by using elementary SPICE diodes and DC voltage sources. Their role is summarized as follows: if $V_r$, i.e. the voltage across the capacitor $C_r$ (which is described by the value of parameter $r(V_M, t)$) falls below $V_1$ (rises above $V_2$) then diode $D_1$ ($D_2$) is forward biased and thus $V_r$ is maintained equal to $V_1$ ($V_2$). In this setup the values of the DC sources are set equal to the boundary values of $r(V_M, t)$; i.e. $V_1 = r_{MIN}$ and $V_2 = r_{MAX}$ [78].

Four different memristor-based circuits were constructed on MULTISIM platform: a simple memristor-resistor circuit (Fig. 5a), memristor-based Wien bridge (Fig. 5b) and a bridge synapse (Fig. 5c). It has been shown, that memristive systems, and especially memristor-based synapse bridge, possess great efficiencies at generating higher harmonic components [79]. The latter was also incorporated in a feedback loop reservoir device: a single node echo state machine (cf. Fig. 6).

The performance of the circuits was initially tested with sine wave signals of 55, 110 and 220 Hz frequencies (referred to musical notes $A_1$, $A_2$ and $A_3$, as named in Scientific/International Pitch Notation [80]). Subsequently various musical intervals (ranging from unison up to octave) were constructed in C-major scale spanning ca. 2½ octaves in natural tuning (Fig. 7). Each tone was represented by a single sine wave pulse of given frequency lasting 2 s with a dumping factor of 2 s$^{-1}$. Signal dumping was necessary to avoid effects associated with abrupt pulse truncation.

Table 1. A memristor description in PSSpice syntax according to [78].

```
.SUBCKT mem1 plus minus

.SYNTAX PSspice
*Parameters' values
.param rmin = 100
.param rmax = 390
.param rinit = 390
.param alpha = 40000
.param beta = 10
.param gamma = 0.2
.param VtR = 1.5
.param VtL = -1.5
.param yo=0.0001
.param m = 82
.param fo = 310
.param Lo = 4
.param Dbreak = 1
.param Dbreak = 1

Gr1 0 r value = {dr_dt(V(plus)-V(minus))*st_f(-(V(plus)-
V(minus)))}
Gr2 0 r value = {dr_dt(V(plus)-V(minus))*st_f(V(plus)-
V(minus))}

D1 k r {Dbreak}
V1 k 0 {rmin}
D2 r g {Dbreak}
V2 g 0 {rmax}
Cr r 0 1 IC={rinit}

*Current equation Imem = V / R(L)
Gpm plus minus value={(V(plus)-
V(minus))/((fo*exp(2*L(V(r))))/L(V(r)))}

*Func. for non-linear threshold-based behavior
.func dr_dt(y)={-alpha*((y-VtL)/(gamma+abs(y-VtL)))*st_f(-
y+VtL)-beta*y*st_f(y-VtL)*
+st_f(-y+VtR)-alpha*((y-VtR)/(gamma+abs(y-VtR)))*st_f(y-VtR)}

*smoothing function
.func st_f(y)={1/(exp(-y/yo)+1)}
*L(V) function
.func L(y)={Lo-Lo*m/y}

.ends mem1
```

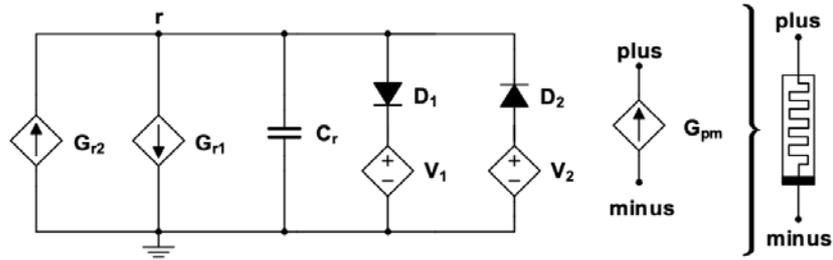

*Figure 4. An equivalent circuit of the memristor model used in this study. Adapted from Ref. [78].*

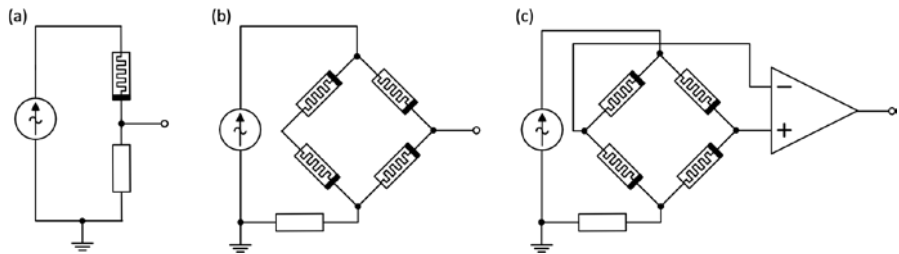

*Figure 5. Memristor-based circuits implemented on the MULTISIM platform.*

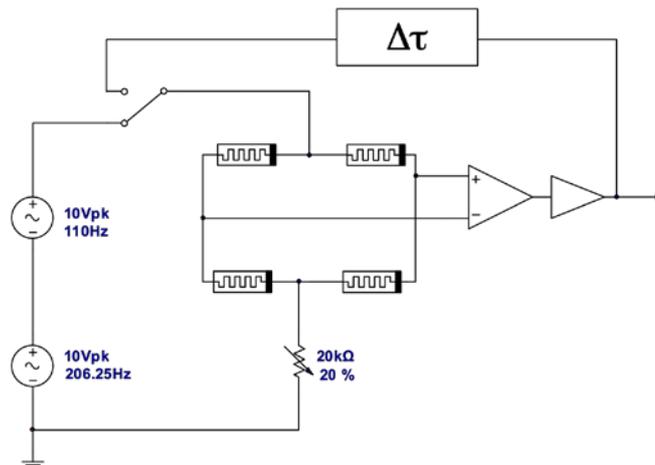

*Figure 6. Single-node echo state machine (SNESM) implemented on the MULTISIM platform.*

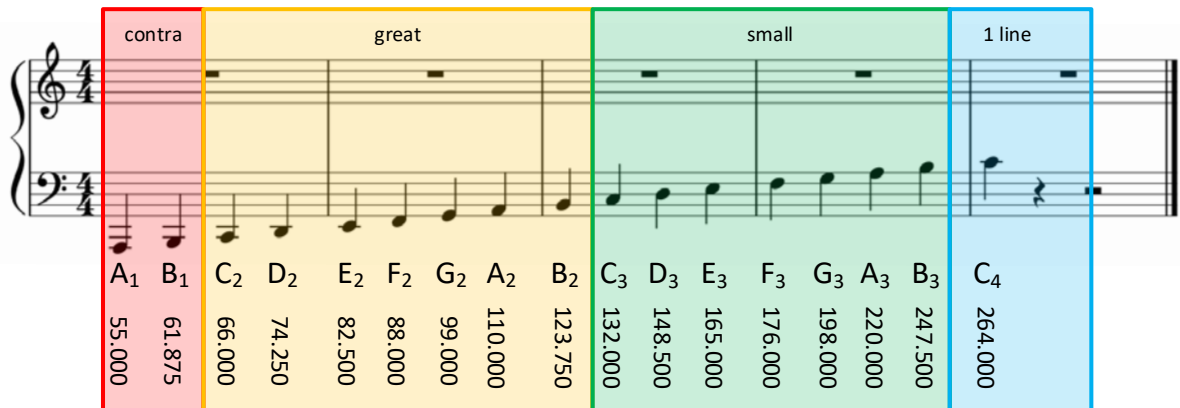

*Figure 7. Sounds of the C-major scale used in this paper, frequencies of each tone are calculated according to the natural tuning with the reference tone $A_4$ tuned to 440 Hz.*

Each simulation was run for 20 s, which allows recording on an input and 9 echoes generated by the feedback loop (Fig. 8). Recorded signals were imported to OriginPro2019 software package for further processing. Fourier transform spectra were calculated using the rectangle window function independently for each signal package.

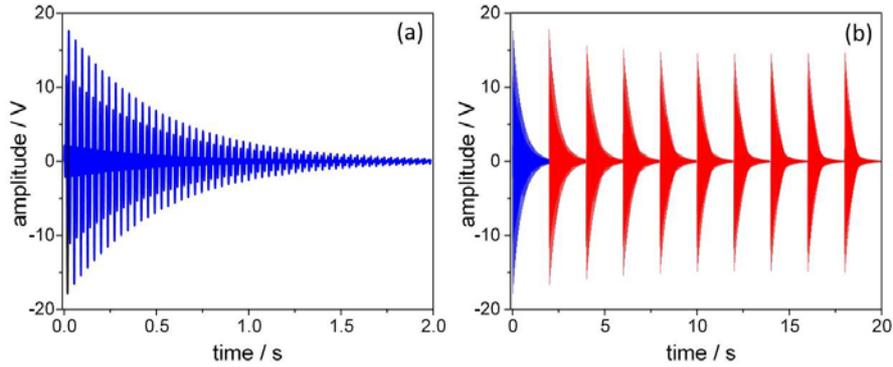

*Figure 8. An example of the input (a) and output (b) signals recorded for the reservoir feedback loop: input perfect fifth ($A_1$-$E_1$, 55 and 82.5 Hz, blue) and output echoes (red).*

**Results and discussion**

Initial tests have involved all model circuits, on the basis of which the most promising circuit has been selected for construction of a single node echo state machine. A single sine input signal of $A_1$=55 Hz frequency has been selected as a test signal. The response of all test circuits has been recorded and compared according to two factors: (i) the depth of the hysteresis curve, i.e. the difference in current between the high and low resistive states upon dynamic stimulation and (ii) the richness of the Fourier spectrum of the resulting spectrum, i.e. the total harmonic distortion. All three circuits (Fig. 5) provide a significant signal transformation, however each of them presents distinct features, which are used to select the most promising circuit for further investigations.

A simple memristive circuit comprising one memristor and one resistor of 2 kΩ (Fig. 5a) yields a classical, significantly asymmetric pinched hysteresis loop (Fig. 9a). The surface areas of lobes are: 54.99 a.u.$^2$ (positive lobe) and 37.06 a.u.$^2$ (negative lobe). The asymmetry of both lobes is a results of numerical instability of the model circuit at low frequencies in MULTISIM environment. Fourier analysis indicates the presence of progressive formation of higher harmonics, the intensity of which decreases exponentially with increasing input frequency (Fig. 9b). Whereas the generation of a complex Fourier pattern is satisfactory and should be beneficial for dissonance/consonance discrimination, the asymmetry of the lobes may bring additional, undesired distortions of symmetric sine signals. Therefore, more complex circuits have been designed and tested for signal processing.

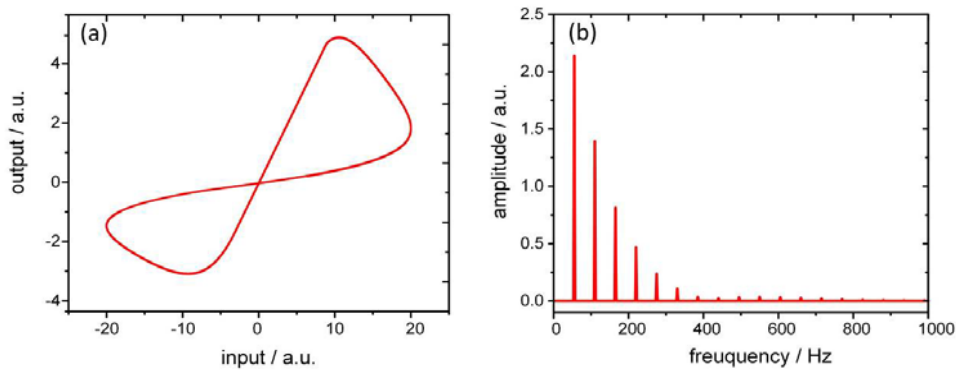

*Figure 9. Pinched hysteresis loop (a) and a Fourier spectrum of an output signal for calculated for a simple memristive circuit from Fig. 5a with input frequency* $A_1$=55 Hz.

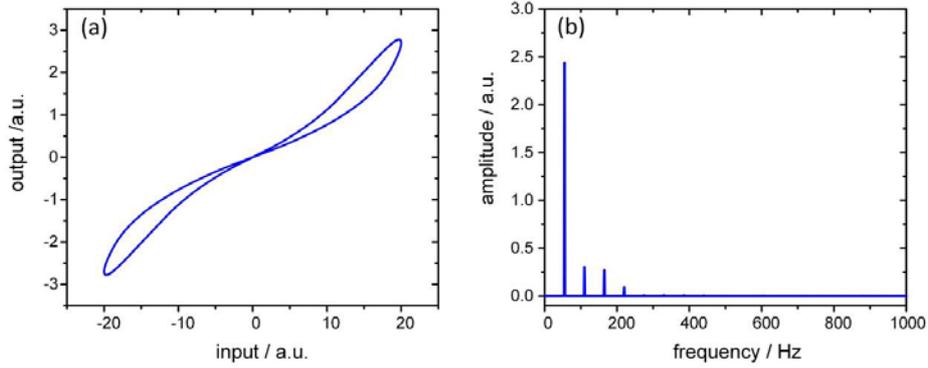

*Figure 10. Pinched hysteresis loop (a) and a Fourier spectrum of an output signal for calculated for a memristor bridge circuit from Fig. 5b with input frequency $A_1$=55 Hz.*

A memristor Wien bridge-like circuit with an additional resistor of 2 kΩ (Fig.5b) generates much less pronounced hysteresis loop, however it has much higher symmetry. Both lobes have exactly the same area of 7.04 a.u.$^2$, which is over 5 times less than in the previous case. The high symmetry observed here is a result of specific connectivity within the bridge. The two memristors in each branch have opposite polarizations, and the two branches are oppositely polarized as well. As a consequence, both branches should have almost the same (within the numerical accuracy) resistance irrespectively of their history. Applied signal induce changes of the resistance ratios in each branch depending on their polarities [81]. The small discrepancies within each branch are reflected in small lobes of the pinched hysteresis loop with lower current values at each memristor, which should result in more reliable results due to elimination of numerical instabilities. Fourier analysis indicates the presence of progressive formation of higher harmonics, however their intensities are very low (Fig. 10b). In comparison with the previous system higher symmetry of hysteresis loop is a significant advantage, however small harmonic distortion (i.e. low intensity of higher Fourier components) should be considered as a drawback.

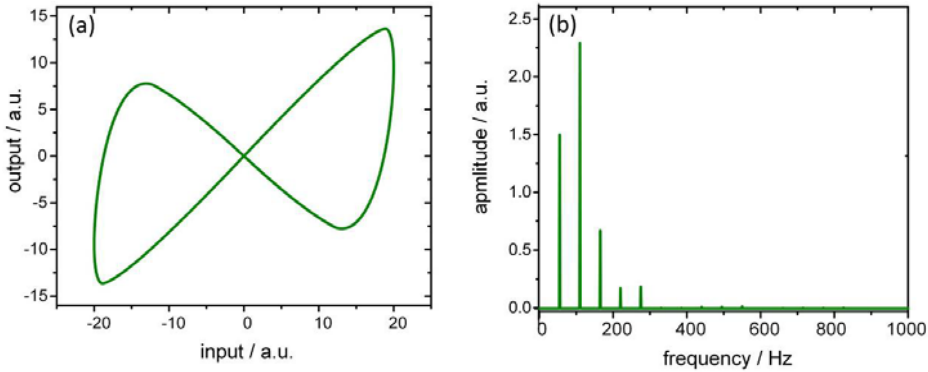

*Figure 11. Pinched hysteresis loop (a) and a Fourier spectrum of an output signal for calculated for a memristor synaptic circuit from Fig. 5c with input frequency $A_1$=55 Hz.*

Incorporation of a differential amplifier to the memristor Wien bridge circuit (Fig. 5c) results in dramatic increase of the hysteresis loop area, and at the same time the symmetry of the pinched loop is preserved (Fig. 11a). The areas of lobes of 239.50 a.u.$^2$ and 239.51 a.u.$^2$ (for 55 Hz/20V input) is the largest one among all tested circuits. This circuit provides also very high harmonic distortion – the intensity of the first harmonic component (110 Hz) is higher than the input signal (Fig. 11b). Therefore, this circuit will be used as a nonlinear node in the single-node echo state machine (SNESM, Fig. 6). The stability of the circuit is a results of reversed-polarity arrangement of memristors in each branch of the bridge. Whereas the total resistance of each branch remains virtually constant, the resistance ratios of upper and lower memristor in each branch change. Therefore, the potential difference between the two apices strongly depends on the history of the system and is also polarity dependent. The high symmetry observed here is a consequence of the mirror symmetry of the bridge. Furthermore, lower current values at each memristor (due to partition of the current within two branches of the bridge) should result in more reliable results due to elimination of numerical instabilities.

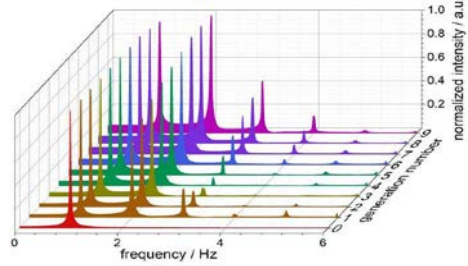

*Figure 12. $A_1$ (55 Hz) input evolution within a single node echo state machine from Fig. 6. Each spectrum corresponds to subsequent step in evolution, called generation. The $0^{th}$ generation is the input signal.*

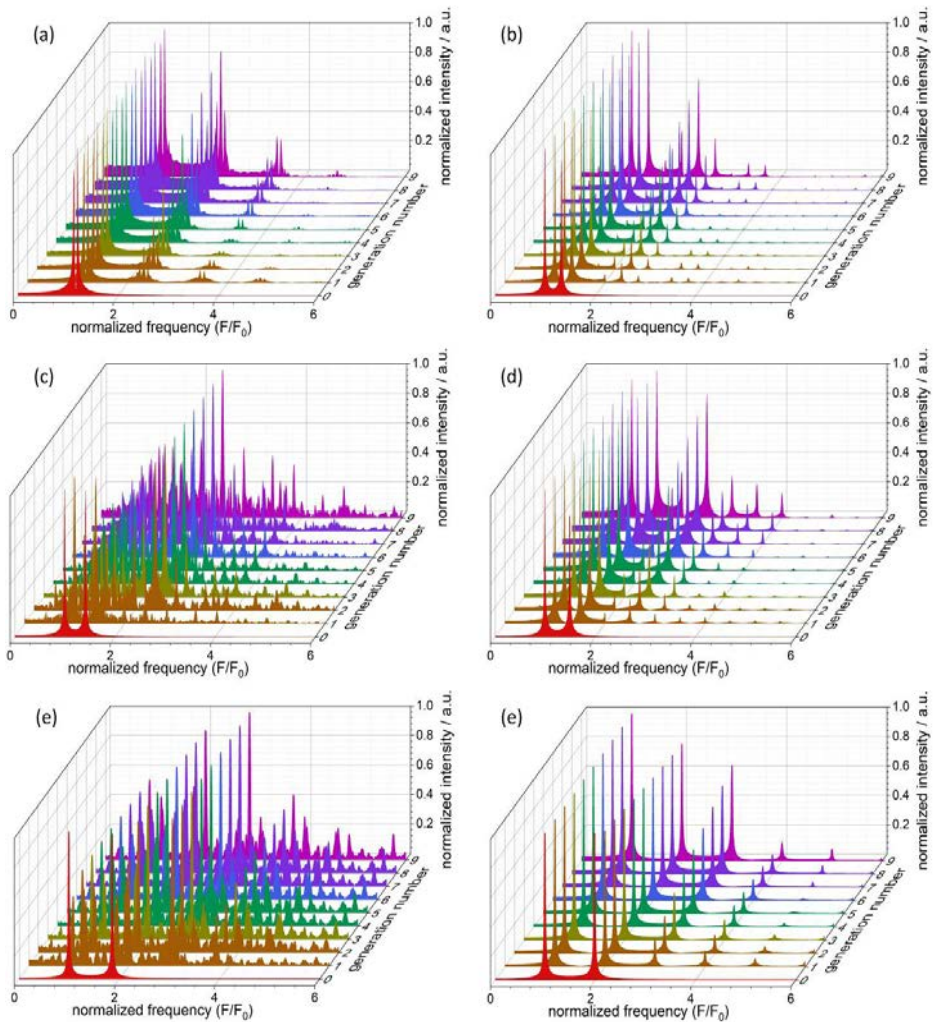

*Figure 13. Evolutions of signals corresponding to selected musical intervals: minor second $B_1$-$C_2$ (a), perfect fourth $C_2$-$F_2$ (b), triton $B_1$-$F_2$ (c), perfect fifth $A_1$-$E_2$ (d), major seventh $C_2$-$B_2$ (e) and octave $A_1$-$A_2$ (f). In each case the $F_0$ frequency is the frequency of lowest tone in the interval.*

*Each spectrum corresponds to subsequent step in evolution, called generation. The $0^{th}$ generation is the input signal. See Fig. 7 for frequency values for each tone.*

AC signals applied as the input to the SNESM circuit (Fig. 6) undergo gradual changes due to a nonlinear character of the memristive node. Single sine signal (*e.g.* 55 Hz, Fig. 12) is transformed within a SNESM circuit into a complex signal with numerous overtones, the intensity of which varies from one generation to the other. Interestingly the ratio of overtone intensities varies among different generations, but the spectral composition of the signal remains unchanged.

More complex behavior can be observed in the case of musical intervals, i.e. signals being a sum of two sine waves of different frequencies applied as the input to the SNESM circuit (Fig. 13). The case of octave is the simplest one. Qualitatively the octave ($A_1$-$A_2$ in this example) is undistinguishable from unison $A_1$. In the case of unison, the first harmonics appears in the first generation signal and slowly increases its intensity. In the case of octave signal the first harmonics is already present in the input, but the higher harmonics appear from the first generation. All spectra contain only integer-value harmonics, *i.e.* $f_0$, $2f_0$, $3f_0$, etc…

Regular patterns of Fourier components, however with different distance between subsequent peaks can be also observed in the case of perfect fourth and perfect fifth. In perfect fifth the intervals between subsequent peaks are identical and equal to ½$f_0$, whereas in the case of perfect fourth ⅓$f_0$. Importantly, the same pattern is stable in all generations, higher harmonics slowly increase with the evolution time. Interestingly, in both cases low frequency components, i.e. ½$f_0$ (in the case of perfect fifth) and ⅓$f_0$ and ⅔$f_0$ can be observed, especially in higher generation signals. This implies that the nonlinear memristive node acts as a frequency mixer, producing both sum and difference frequencies (eqs. 5-6):

$$f_+^1 = f_2 + f_1 \tag{5}$$

$$f_-^1 = f_2 - f_1 \tag{6}$$

Taking into account relations between frequencies in perfect fifth (eq. 7) and perfect fourth (eq. 8) in natural scale [33]:

$$f_2 = \tfrac{3}{2} f_1$$

or (7)

$$f_2 = \tfrac{4}{3} f_1 \tag{8}$$

the observed low frequency signal, as well as high frequency component with non-integer spacings is obvious. In the first generation signal additional components within perfect fifth (eqs. 9-10) and perfect fourth (eqs. 11-12) can be found:

$$f_+^1 = \tfrac{3}{2} f_1 + f_1 = \tfrac{5}{2} f_1 \tag{9}$$

$$f_-^1 = \tfrac{3}{2} f_1 - f_1 = \tfrac{1}{2} f_1 \tag{10}$$

or

$$f_+^1 = \tfrac{4}{3} f_1 + f_1 = \tfrac{7}{3} f_1 \tag{11}$$

$$f_-^1 = \tfrac{4}{3} f_1 - f_1 = \tfrac{1}{3} f_1 \tag{12}$$

In the second generation additional components are formed due to frequency mixing of input tones and the first generation ones. In the case of perfect fifth two new frequencies are equal either input tones or the first generation ones, as well as some 2$^{nd}$ generation frequencies are identical (only combination yielding positive differential frequency are listed, 13-20):

$$f_1^2 = f_1 + f_-^1 = f_1 + \tfrac{1}{2}f_1 = \tfrac{3}{2}f_1 = f_2 \tag{13}$$

$$f_2^2 = f_1 - f_-^1 = f_1 - \tfrac{1}{2}f_1 = \tfrac{1}{2}f_1 = f_-^1 \tag{14}$$

$$f_3^2 = f_1 + f_+^1 = f_1 + \tfrac{5}{2}f_1 = \tfrac{7}{2}f_1 \tag{15}$$

$$f_4^2 = f_2 + f_-^1 = \tfrac{3}{2}f_1 + \tfrac{1}{2}f_1 = \tfrac{4}{2}f_1 \tag{16}$$

$$f_5^2 = f_2 - f_-^1 = \tfrac{3}{2}f_1 - \tfrac{1}{2}f_1 = \tfrac{2}{2}f_1 \tag{17}$$

$$f_6^2 = f_2 + f_+^1 = \tfrac{3}{2}f_1 + \tfrac{5}{2}f_1 = \tfrac{10}{2}f_1 \tag{18}$$

$$f_7^2 = f_+^1 + f_-^1 = \tfrac{5}{2}f_1 + \tfrac{1}{2}f_1 = \tfrac{6}{2}f_1 = f_4^2 \tag{19}$$

$$f_8^2 = f_+^1 - f_-^1 = \tfrac{5}{2}f_1 - \tfrac{1}{2}f_1 = \tfrac{4}{2}f_1 = f_5^2 \tag{20}$$

Numerous iteration of the composite signal in the feedback loop result in accumulation of various frequencies with the spacing of ½$f_0$ (perfect fifth) or ⅓$f_0$ (perfect fourth). Thus, the feedback loop is utilized to recursively process signal and utilize concepts of reservoir computing [74,82,66,83-86] for attempted classification of signals in terms of musical quality.

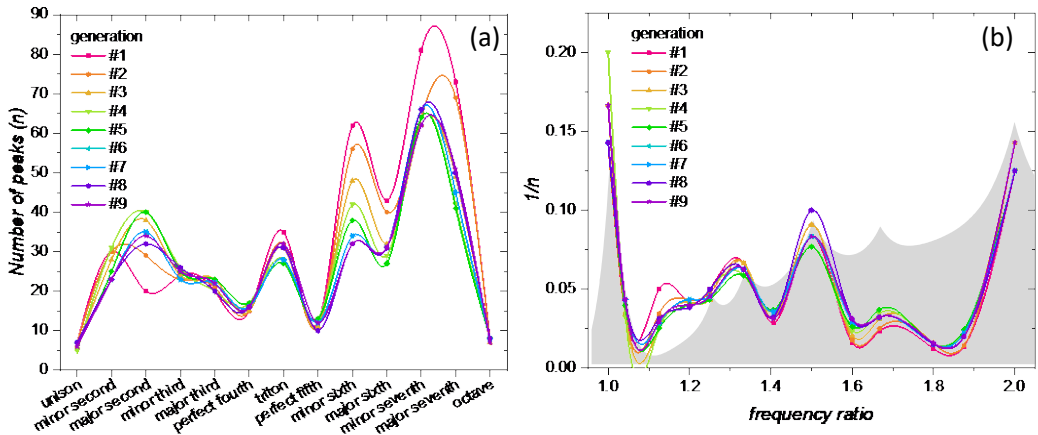

*Figure 14. Number of peaks for various music intervals for subsequent Fourier transforms of the signal circulating in the feedback loop (a) and a reciprocal number of peaks as a function of frequency ratio for various intervals. Area shaded in grey represents simulated "sensory consonance" curve (mirror image of the dissonance curve from Fig. 3c for 5 harmonics). Points represent experimental data, whereas solid lines serve as eye guides. Data presented for three low base frequencies ($B_1$=61.875, $C_2$=66 and $E_2$ = 88 Hz).*

Other intervals generate much more complex patterns within the SNESM circuit and their evolution results in continuously increasing number of Fourier components, which are not evenly spaced. This observation may be already considered as a simple classification criterion: consonant tone combinations result in outputs with relatively small number of spectral components, and these components are evenly spaced. Whereas this is a useful criterion of preliminary classification, but it does not allow rating of degree of dissonance, as is it usually performed by listeners (Fig. 3a). Therefore, two different classification protocols have been suggested to classify musical intervals according to their dissonance/consonance levels and to unambiguously measure the degree of dissonance of a given tone combination: (i) counting a total number of peaks in Fourier spectrum in each generation and (ii) measuring distances between neighboring peaks in Fourier spectrum in each generation. These two approaches are briefly discussed below.

A simple peak count analysis was performed to examine if any patterns were present in the obtained data (SciPy – Python library for scientific computing was used for this task). Detection of peaks was based on searching of the local maxima in a given point range by simple comparison of neighboring values. It was decided that the peaks should consist of at least 4 points from base to base as an arbitrary "peak" criterion. This type of peak detection was carried out for all Fourier transforms among all delayed feedback loop repetitions. For each interval, the above analysis was performed for three different base frequencies. The results are summarized for individual intervals as peak count in function of repetition of the feedback loop in order to examine data for deviations and anomalies. In addition, obtained number of peaks was summarized in the function of individual intervals for each generation.

Whereas initial input (generation #0) contains always two Fourier components, the higher generations contain variable number of peaks, moreover it also evolves in time, as expected from recurrent character of the SNESM circuit and Equations (9)-(11) as shown in Fig. 14a. It can be noticed that generally the number of peaks for consonant intervals (unison, perfect fourth, perfect fifth and octave) are significantly lower than for the dissonant intervals. Therefore, in order to reproduce the sensory dissonance curve (cf. Fig. 3), the reciprocal number of Fourier components was plotted against the frequency ratios of various intervals (Fig. 14b). The line connecting obtained points was added to guide the eye (spline in OriginLab Pro). Surprisingly, it resembles the Helmholtz curve of consonances and dissonances, with exception of minor and major thirds, which form a local minimum on experimental curve, but are usually recognized as weak consonances (cf. Fig. 3). The other discrepancy is minor sixth, which, along with major sixth constitutes a local minimum, as is again recognized as a weak consonance. Despite this small discrepancies, the reservoir system with a single memristor bridge synapse shows unexpectedly good performance in discrimination of musical dissonance.

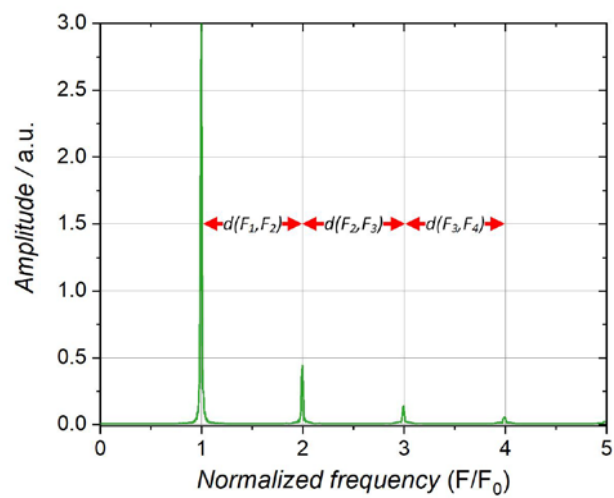

*Figure 15. A principle of signal analysis based on distances between peaks in Fourier spectra.*

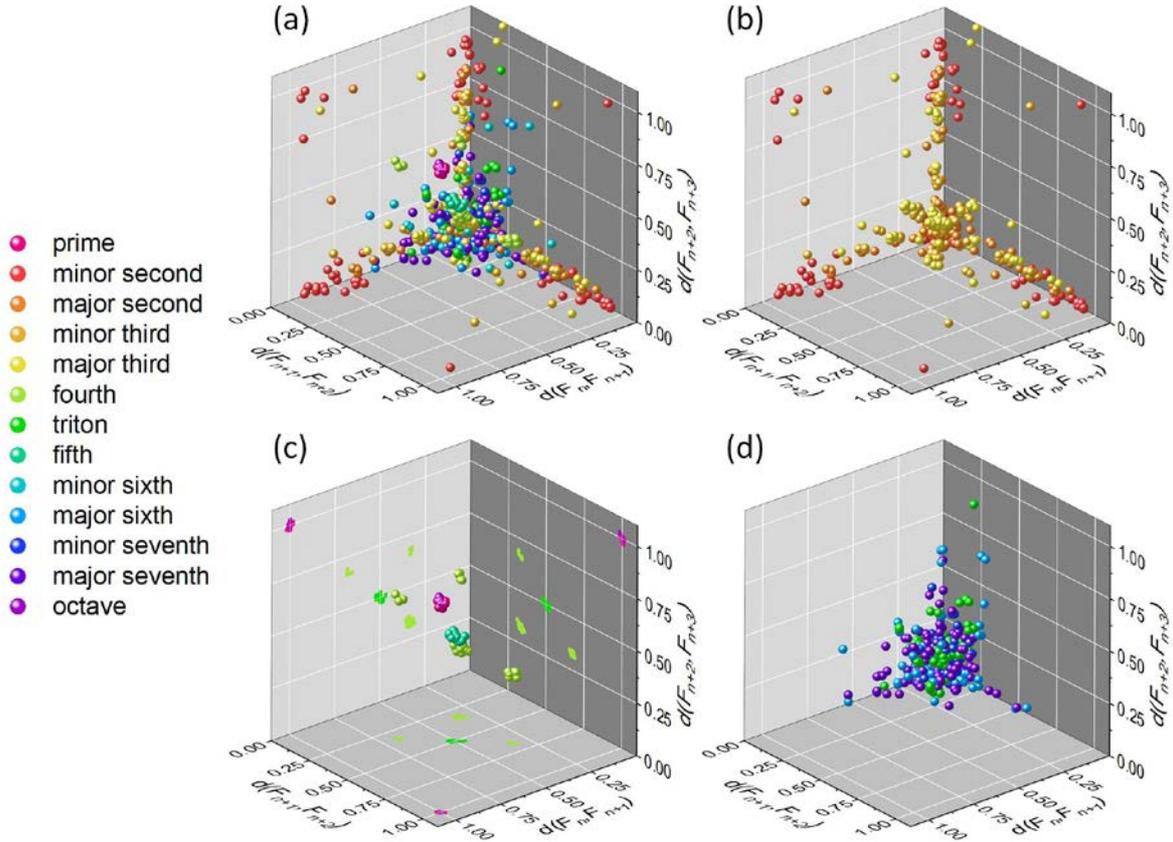

*Figure 16. Geometrical representation of various intervals on the basis of distances between subsequent components in Fourier spectra. All intervals (a), seconds and thirds (b), unisons, octaves, fourths and fifths (c) and tritons, sixths and sevenths (d).*

In order to examine the relationship between the normalized frequencies of the peaks present in the Fourier transforms, an analysis was performed using the difference of normalized frequencies between a given peak and the peak preceding it (Fig 15). Then to visualize and analyze obtained data, a three-dimensional "phase portrait" was constructed using the obtained ⅔ frequency differences. For each interval and for each peak (denoted here as $F_n$) a set of distances to the nearest neighbors: $d(F_n, F_{n+1})$, $d(F_{n+1}, F_{n+2})$ and $d(F_{n+2}, F_{n+3})$ was calculated. These values were used as coordinates defining points in space for all subsequent peaks present in a given spectrum. In the data representation, the given point is obtained on the basis of given difference and two successive frequency differences. Each interval is thus represented as a set of points in three dimensional space (Fig. 16). In order to enrich the constructed phase portrait all signal generations – the whole history of signal evolution – have been employed. Due to this method of data visualization, the intervals possessing greater number of peaks in the Fourier spectra are represented by a greater number of points in the obtained phase space (cardinality of their set is greater). It can be observed, that intervals regarded as absolute – unison and octave – are clustered in one place, around the (1,1,1) point, which is obvious, because the differences between the peaks of normalized frequencies for these intervals will always be 1. In turn, the intervals treated as perfect – fourth and fifth – are clustered in separation from other intervals. Perfect fourth is clustered in 4 different regions around points (⅓,⅓,⅓), (⅓,⅓,⅔), (⅓,⅔,⅓) and (⅓,⅔,⅔). Perfect fifth can be found in one bigger cluster at (½,½,½). This observation is consistent with the previous analysis of the distance of generated harmonics. Only points corresponding to these consonant intervals are clustered in very limited spaces. Other intervals, usually considered dissonant, are localized in different places. Seconds and thirds occupy places along axes with a cluster close to the origin of the coordinate system. Triton, sixths and sevenths together are scattered as well close to the origin of the coordinate system. Most intervals perceived as dissonant can be found in one big cluster of points near the center of the graph. Apart from absolute and perfect intervals, the other intervals are usually clustered in several separate areas, but their points can also be found mixed with other intervals in different parts of the graph.

The two different classification schemes based of various processing of reservoir output present very similar results. The approach based on peak counting yields classification closely related to the auditory classification of musical intervals. The geometrical approach, based on distances between subsequent peaks in Fourier spectra clearly differentiates consonances and dissonances. Furthermore, it classifies consonances into two categories (ideal – unison ad octave and perfect – fourth and fifth). In the case of dissonant intervals, this method also provides some kind of differentiation, in which seconds and thirds are, at least partially, separated from sixths, sevenths and triton.

**Conclusions**

Two computation examples presented in this paper show that a simple reservoir computing system – a single node echo state machine with appropriately designed nonlinear node: a memristive bridge synapse with differential amplifier, is capable of advanced signal processing and classification of musical intervals according to their harmonic quality. The system generates a family of higher frequency components in the case of single sine input, and a series of differential frequencies in the case of two sine inputs. Analysis of harmonic components showed the highest performance of the selected non-linear element among other similar variants for the computational node. Surprisingly, the results are very similar to the evaluation of sound samples by human subjects (Fig. 3a) and theoretical dissonance analysis using the Plomp-Levelt [38] approach and Sethares algorithm [33] .

Classification of musical intervals is performed on the basis of two simple post-processing protocols (readout layer), depending on the analysis of the obtained Fourier spectra. By simple peak counting and measuring distance between them it is possible to reproduce the sensory dissonance curve and also to classify intervals in three-dimensional space according to the spectral characteristics of signal processed in the reservoir. In the obtained phase-space it is possible to separate absolute intervals (unison and octave), individual perfect intervals (fourth and fifth) and dissonant intervals (and to some extent seconds and thirds). Interestingly, this classification yields results very similar to other numerical models, but in this case completely devoid of any preliminary bias/training based on theory of music, in a purely unsupervised manner.

In typical musical context (with an exception of electronic music) single sine waves are very uncommon. As it can be seen from the numerical dissonance analysis (eq. 1-4, Fig. 3), the notion of consonance and dissonance makes sense only in the case of higher harmonics: the interval of octave becomes distinguishable from the whole acoustic spectrum with at least one harmonics, the perfect fifth requires at least two harmonics. Interestingly, the system presented here can deal with single sine waves and classifies the notes in a similar way as human subjects and established numerical algorithms. The presented data demonstrate the power of reservoir computing in solving difficult tasks. These results also suggest, that neural processes involved in perception of music may be related to the reservoir computing principles.

**Acknowledgements**

Authors acknowledge the financial support from the Polish National Science Centre within the MAESTRO project (grant agreement No. UMO-2015/18/A/ST4/00058). DP has been partly supported by the EU Project POWR.03.02.00-00-I004/16.

**Data availability**

The data that support the findings of this study are available from the corresponding author upon reasonable request. All recorded time series, Fourier spectra and other computational results are available as asci as well as Origin-specific binary files.

**Conflicts of interest/Competing interests**

Authors declare no conflicts of interest.

**Code availability**

No special codes have been developed, all computations were done either with commercial software packages (Multisim, OriginPro) or with freely available Python libraries. The modified QBasic code (based on original code by W.A. Sethares and modified to the QBasic syntax) for dissonance calculation is available from the authors on request.

**Author contributions**

Dawid Przyczyna has conducted most data analysis and has written large fragments of the manuscript. Maria Szaciłowska has performed most of numerical experiments in Multisim. Marcin Strzelecki has prepared sections of musical context of the research, contributed in the discussion of the results and revised the whole manuscript. Marek Przybylski has contributed in the discussion of results and has partially revised the manuscript. Konrad Szaciłowski has formulated the leading idea, has designed the Multisim calculation schemes, performed most of the raw data processing and has written large sections of the manuscript.